# Shear band relaxation in a deformed bulk metallic glass

I. Binkowski[1], G. P. Shrivastav[2], J. Horbach[2], S.V. Divinski[1], G. Wilde[1]

[1]*Institute of Materials Physics, University of Münster, Wilhelm-Klemm-Str. 10, 48149 Münster, Germany*

[2]*Institute of Theoretical Physics II, University of Düsseldorf, Universitätsstr. 1, 40225 Düsseldorf, Germany*

Relaxation of shear bands in a $Pd_{40}Ni_{40}P_{20}$ bulk metallic glass was investigated by a combination of radiotracer diffusion and molecular dynamics (MD) simulations, allowing to determine for the first time the effective activation enthalpy of diffusion along shear bands in a deformed glass. The shear bands relax during annealing below the glass transition temperature and the diffusion enhancement reveals unexpectedly a non-monotonous behavior. The development of shear bands and the subsequent relaxation of stresses after switching off the shearing are characterized on microscopic to mesoscopic length scales by MD simulation subjecting the model glass to a constant strain rate. Mean-squared displacements as well as strain maps indicate that the heterogeneity, as manifested by shear bands in the systems under shear, persist after the shear is switched off. We observe a further relaxation of residual stresses that remain localized in regions where the shear band has been present before, although the system is – different from the macroscopic experiment – homogeneous with respect to the local density. These results indicate that even on a local scale one may expect strong dynamic heterogeneity in deformed glassy solids due to shear banding. The results thus suggest that plastically deformed metallic glasses present poly-amorphous systems that necessitate descriptions that are analogous to multiphase materials including the presence of heterophase interfaces.

Although non-homogenous plastic deformation of bulk metallic glasses (BMGs) via the formation of shear bands attracted increased attention in the past [1, 2], it is still far from being resolved, see e.g. the reviews in [3, 4]. As a generally accepted concept, so-called "shear transformation zones" (STZ), i.e. areas in which groups of atoms collectively undergo a local shear transformation, have been introduced [5] as "unit carriers" of plastic deformation in metallic glasses. A cross-over from random 3-dimensional shear events (STZ formation) to correlated 2-dimensional dynamics has been brought forward to explain the observed shear banding in metallic glasses [6]. Although the activation of a single STZ event is not inevitably related to a change of the excess volume, shear bands are often described in terms of excess volume accumulation [3, 7]. Recently, we investigated diffusion in deformed $Pd_{40}Ni_{40}P_{20}$ (at. %; in what follows we will use the abbreviation PdNiP) glass, in which almost a single family of shear bands was introduced, and were able to unambiguously prescribe the observed enormous enhancement of the diffusion rate to an ultra-fast atomic transport along these quasi-2-dimensional pathways [8]. The tracer concentrations in the corresponding concentration profiles, which were related to shear band diffusion, were shown to scale with the number density of the introduced shear bands (that in turn scales with the imposed strain), substantiating the reliability of the measurements. Nevertheless, dedicated TEM measurements revealed strong and probably characteristic changes of the specific volume along shear bands [9, 10], with alternating regions of diluted and denser regions.

These results indicate significant modifications of the local structure of the glassy material inside the shear bands compared to the surrounding matrix, which have been revealed by local analyses via fluctuation electron microscopy [9, 10], too. Additionally, these results point out that deformed glasses containing shear bands might be treated in analogy to a two-phase material including heterophase interfaces, with the second phase being the shear band phase that has undergone strong structural modifications due to the applied shear. Such structural modifications should also affect characteristic glass properties such as the relaxation dynamics and, potentially, might even affect the structure and properties of the matrix zones surrounding the shear bands via an exchange of the excess volume. On the other hand, structural relaxation is a characteristic process of glasses and thus serves as a sensitive probe for analyzing the presence of such an apparent poly-amorphicity in deformed metallic glasses.

In order to address the relaxation behavior of shear bands, in the present paper the diffusion rates in plastically deformed PdNiP glasses are experimentally measured as function of temperature and varying annealing and pre-annealing times. These measurements target the relaxation behavior that is specific of the glassy material inside the shear bands only, which is not possible by macroscopically averaging measurements, since the fraction of matter inside the shear bands is only of the order of $10^{-4}$ in a moderately deformed glass. Such information is, however, vital in order to understand the modifications of the structure of the glass inside the shear bands, as well as the possible coupling of excess volume fluxes and excess volume re-distribution inside the shear bands and within the surrounding matrix.

Complementary to the experiments, shear band relaxation is investigated by molecular dynamics (MD) computer simulations of a simple model of $Ni_{80}P_{20}$, namely a binary glass-forming Lennard-Jones (LJ) mixture, sheared with a constant strain rate. A small shear rate is applied that is associated with the formation of a shear band. Subsequently, the stress relaxation was analyzed by switching off the shear at different strain values. As a matter of fact, the shear stress does not decay to zero after switching off the shear field, but it tends to approach a finite value in the long-time limit and thus the resulting glass structure, albeit not sheared anymore, is established to remain under stress. Different from the response in the supercooled liquid state [11], such residual stress has been found to be the generic case with respect to stress relaxation after switching-off of shear in glassy solids [12]. Furthermore, the residual stresses are found to be *heterogeneously* distributed in the glass sample, thus reflecting the memory of the shear band, as formed in the sheared system. We analyze the formation of shear bands by a methodology that we have recently developed in the context of a sheared soft-sphere mixture in its glass state [13, 14]. To this end, maps of the local mean-squared displacement (MSD) are computed.

The simulations show that on microscopic to mesoscopic length scales the shear bands do not lead to heterogeneities that can be characterized by a significant variation of the local density, as found experimentally on significantly larger scales [9, 10]. However, residual stresses are spatially localized in the region where the shear band was created earlier and the simulation can disentangle how these localized residual stresses affect the mechanical properties of the glass. Moreover, the direct comparison between experimental results (which involve macroscopically averaged data) and simulations allows analyzing the unexpectedly complex relaxation behavior of shear bands in metallic glasses that needs to be taken into account for any constitutive description of the plastic response of metallic glasses.

## Shear band relaxation as revealed by tracer diffusion

Four temperatures were chosen for the present study and the measured concentration profiles reveal unambiguously a striking diffusion enhancement related to the presence of shear bands in the deformed glass (see supplementary information where the parameters of diffusion annealing treatments and the determined profiles are shown). The determined effective diffusion coefficients are plotted in Fig. 1a as function of the inverse temperature.

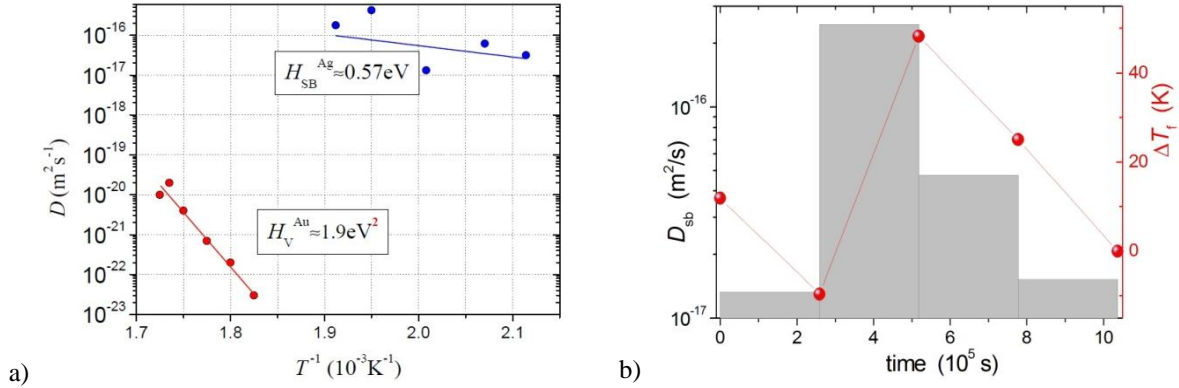

**Figure 1: Shear band diffusion in deformed PdNiP glass. a**, The Arrhenius plot for $^{110m}$Ag diffusion along shear bands (blue symbols) in comparison to the diffusion coefficients measured for Au bulk diffusion in glass of nominally the same composition (red symbols). **b**, The shear band diffusivity, $D_{gb}$, (gray bars, left ordinate) and the difference in the fictive temperatures, $\Delta T_f = T_f^{def} - T_f^q$, of deformed, $T_f^{def}$, and as-quenched, $T_f^q$, glasses (red spheres, right ordinate) as functions of the total annealing time at 498 K.

The diffusion coefficients associated with Ag diffusion in shear bands follow an Arrhenius line and an effective activation enthalpy of about (55±10) kJ/mol is determined from the corresponding slope. Since bulk diffusion of Ag in the PdNiP glass has not been measured yet, in Fig. 1a the shear band diffusivity of Ag atoms is compared to the diffusion coefficients of Au in nominally the same glass measured by Duine *et al*. [15]. One recognizes that the effective activation enthalpy of shear band diffusion, $Q_{sb}$, is about one third of that value for bulk diffusion in the glass matrix, $Q_v$, i.e. $Q_{sb}/Q_v \approx$ 55 kJ mol$^{-1}$ / 182 kJ mol$^{-1}$ $\approx$ 0.3. In crystalline solids such ratio would be characteristic for the relationship between the activation enthalpies of surface and bulk diffusion [16]. It should be pointed out here that the specific experimental set-up with application of two tracers (see supplementary information) and the dedicated TEM analyses did not reveal any open and percolating porosity in the deformed specimens, so that the diffusivities measured here represent solid-state diffusion through the shear bands.

The relaxation behavior of shear bands at 498 K is investigated applying the radiotracer diffusion measurements as a sensitive probe of the shear band state. Unexpectedly, a non-monotonous time-dependence of the measured diffusion enhancement is found (Fig. 1b), with the diffusion rate along shear bands first increasing and then decreasing at longer total annealing times.

Simultaneously, we followed structural changes of the glass by Differential Scanning Calorimetry (DSC) and the characteristic changes of heat release at the glass transition temperature were recorded, Fig. 2. As a result of plastic deformation, an explicit "peak" at the glass transition, around 600 K, appears which is absent in the un-deformed state. The peak in the apparent specific heat that is directly proportional to the heat flow measured by DSC is directly related to the relaxation state of the glass [17] and it shows a distinct evolution with annealing time that is markedly different from the un-deformed state. After 3 days of annealing, the glass transition peak appears as two separate maxima, indicating the presence of two regions with different relaxational states (different fictive temperatures [18, 19]). This observation indicates the presence of at least two regions with different excess volume content or basically different local configurations that occupy similar volume fractions, since this signal is observed in a macroscopically averaging measurement.

For a Pd-Ni-P glass of slightly different composition, similar modifications of the glass transition signal have been observed as a function of thermal annealing in the temperature interval of the glass transition [20]. Those modifications of the glass transition signal, that were found to be reversible, but with a significant hysteresis between formation and annihilation temperatures, have been shown to be correlated with modifications of the medium range order. The present findings correlate also with the existence of a broad spectrum of atomic-packing motifs in a bulk metallic glass upon deformation, as recently reported in [21] and the basically different relaxation of boson peaks in deformed glasses with respect to the undeformed state [22, 23], which also indicates strongly that deformation can adjust new local medium range order motifs that, in the view of energy landscapes, might present local minima within a different metabasin. The fact that qualitatively similar observations were obtained for two different glasses (Pd-based [22] or Zr- based [23]) that had been subjected to different types and amounts of strain substantiates the general nature of these findings.

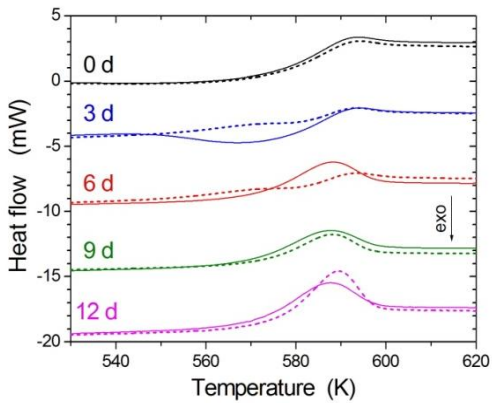

**Figure 2: DSC data**. DSC scans of PdNiP glass in as-quenched (solid lines) and deformed (dashed lines) states without pre-annealing (black lines) and after isothermal annealing treatments at 498 K for 3 (blue), 6 (red), 9 (green) and 12 (magenta) days. Heat flow signals as measured near glass transition temperature at about 580 K are shown (for full DSC scans see supplementary information). The DSC curves for different annealing times are successively shifted along the Y-axis by −5 mW for a better visualization.

Upon continued annealing, the splitting of the peak vanishes, indicating that the majority of the volume of the specimen is in a similar relaxational state again, however fundamentally different from that of the un-deformed glass which does not reveal such a peak, Fig. 2. After 6 days of annealing, the peak at the glass transition is most pronounced and decreases again in height with further annealing. Such complex time dependence of the relaxation peak at the glass transition is indicative of so-called "crossover" behavior [24, 25], where the time scale of the annealing treatment and the time scales for excess volume re-distribution and annihilation compete.

We propose to characterize the deformation-induced changes of the glassy structures by the *difference* between the fictive temperatures [26] of deformed, $T_f^{def}$, and as-quenched, $T_f^q$, glasses, $\Delta T_f = T_f^{def} - T_f^q$. In Fig. 1b, this difference is plotted as a function of the annealing time and is compared to the measured annealing-time dependence of the shear band diffusion coefficient. Since the diffusion annealing time was 3 days in each experiment and since the shear band structure evolves further by relaxation during these heat treatments, the determined effective diffusion coefficients are plotted as bars spanning the time intervals of the given diffusion annealing.

A generally unexpected and striking feature is prominent, the shear band diffusivity evolves non-monotonously with annealing time - it is first increased and then decreased and the annealing time dependence of $\Delta T_f$ is very similar, Fig. 1b. In a simple free-volume picture, this observation is in agreement with the annealing-time evolution of the relaxation peak at the glass transition: the corresponding enthalpy is maximum when the relaxational state has acquired the highest amount of excess volume (corresponding to the highest peak maximum at the glass transition). Figure 1b is the first example of a cross-over behavior observed for shear band diffusion and the fictive temperature excess related to the shear localization almost 200 K below the glass transition temperature. Previously, such a cross-over has been observed, e.g., for Young's modulus [27], Curie temperature [28], or viscosity [25] of metallic glasses far from the glass transition. The distinct feature is that in the present study a metallic glass has undergone a plastic deformation at room temperature instead of equilibration annealing below the glass transition as it was performed in previous investigations.

**Atomistic insights into shear band relaxation dynamics**

In the MD simulation, the glass sample is first sheared applying a constant shear rate of $\dot{\gamma} = 10^{-4}$. As shown recently [29], at this shear rate a shear band forms at a strain of about 0.2, the width of which is growing diffusively with time. Thus, at a certain intermediate value of the strain the system exhibits an inhomogeneous flow pattern with non-flowing regions and regions of high mobility where the shear band is located. The central question that we address here is about the shear stress relaxation after the shear field has been switched off at a certain level of strain. We determine the shear stress as a function of time and analyze to what extent the inhomogeneous flow patterns, as formed in the sheared system, are reflected in a spatially inhomogeneous stress relaxation. Note that the stress is computed from the particle trajectories via the virial formula as in Ref. [11].

Stress relaxation after the given levels of deformation was analyzed, Fig. 3a (for details see the supplementary information). It was found that the stress remains first constant over about half an order of magnitude in time, then it starts to decrease following a power law and approaches finally a constant non-zero value, Fig. 3a. Thus, as a result a glass state subjected to a finite residual stress is obtained.

To obtain information about spatial heterogeneities of the deformed samples with these residual stresses, we analyzed them in terms of maps of the mean-squared displacement (MSD) and the non-affine strain, Figs. 3 and 4. Before the switching-off of the shear field, the MSD in the region where the shear band is located is much higher than in the rest of the system where the system has essentially not undergone the transition towards plastic flow, see Fig. 3b for the case of a sample with 100% shear deformation. In view of tracer diffusion data, one may expect that after the switching-off the mobility of particles is higher in regions where the shear band appeared before. To analyze this effect, we decompose the system into 8 layers along the $z$ direction (cf. illustration in Fig. 3d-f) and compute the average MSD in each of these layers. It is obvious that the MSDs in the shear band region are significantly higher than those in nearby layers (not affected by shear localization).

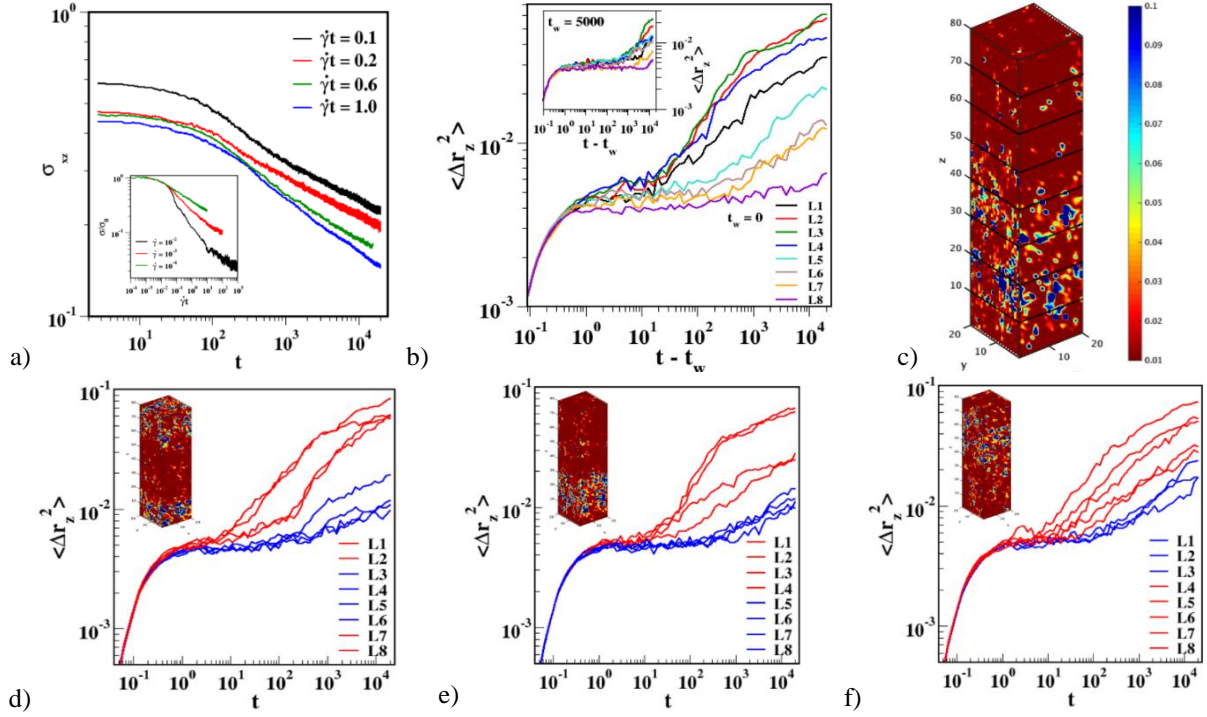

**Figure 3: Atomistic simulation of stress relaxation**. **a**, Stress relaxation after 10%, 20%, 60% and 100% deformation for strain rate $10^{-4}$. The inset shows the stress relaxation for strain rates $10^{-2}$, $10^{-3}$, $10^{-4}$. **b**, Layer-wise mean squared displacement (MSD) of particles during the stress relaxation after 100% deformation. Time origin in (b) is taken at the time when shear is switched off, $t_w = 0$. The inset shows the MSD with time origin shifted at $t_w = 5000$. The color-coded lines correspond to the layers indicated in the snapshot. **c**, Snapshot of system divided into 8 layers of size $10\sigma_{AA}$. **d**, **e** and **f** shows layer-wise MSD for three different samples, layers in the shear band region are shown in red and layers outside shear band are shown in blue.

Recently, evidence has been given that there is a direct link between the macroscopic shear stress and the MSD, reflecting the response of the system to a shear field on a local level [30]. Therefore, one may expect that the behavior of MSD maps is directly linked to the residual stresses in the system; in particular, heterogeneities in the MSD may directly display an inhomogeneous distribution of residual stresses in the system. In Fig. 3b, the time where the shear field has been switched off, $t_0$, is chosen as the time origin for the MSD; the inset displays the corresponding curves for a time origin shifted by $t = 5000$ with respect to $t_0$. One can clearly infer from the figure that at positions where the shear band was located before the particles have still a much higher mobility. In the non-mobile regions, the MSD is essentially constant after the initial microscopic regime indicating the localization of the particles in the glass sample. The same behavior is also evident from the curves in the inset where the time origin is shifted with respect to $t_0$. Also in this case, the MSD increases with respect to the layers where the shear band was located.

Thus, a striking feature of the present simulation is that the diffusion enhancement is highly localized towards a shear band, as it is circumstantiated from the present macroscopic diffusion measurements. Moreover, the memory about the location of the shear band in the deformed sample is not only manifested in the MSD but also in terms of a localization of the non-affine strains in the system. This is indicated by MSD and strain maps that we show in Fig. 4 for the two times $t = 5000$ and $t = 10000$ after the switch off of the shear field. In the shear-band region both the MSD and the strain are significantly higher than in the "inactive" part of the system. From this, it is also evident that the residual stresses are located in the region where the shear band was located before.

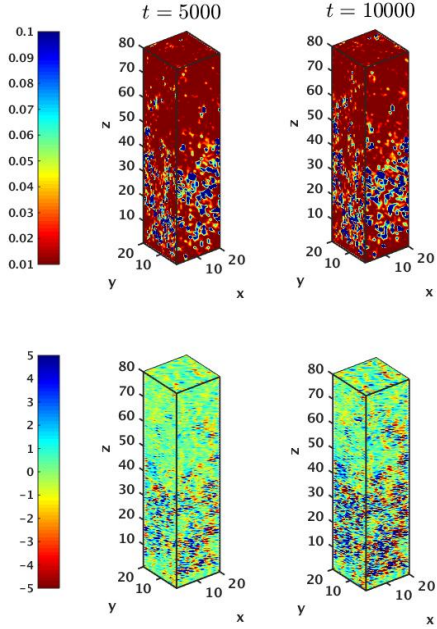

**Figure 4. MSD and strain maps after shear to 100% deformation.** Upper panels show the MSD maps at $t = 5000$ and 10000. Shear is switched off after 100% deformation. Lower panel shows the strain map at times shown in the upper panel.

**Heterogeneity of relaxation behavior of the deformed glass**

A phenomenological model of shear band generation in terms of STZs formation and excess volume accumulation, which is conform to the present findings with respect to kinetic, thermodynamic and atomistic simulation data, is formulated in the supplementary information. The main finding is that the residual stress is frozen in the deformed glass and remains highly localized to shear bands which promoted the imposed plastic deformation. The structure and kinetic heterogeneity is revealed on different levels, starting from the atomistic one (MSD and non-affine strain maps), mesoscopic (stick-slip behavior as found on a sub-micrometer scale using Digital Image Correlation analysis [31]) and the macroscopic level (the present results on tracer diffusion).

The time dependence of the MSD indicates different regimes, with a distinct change at longer times, revealing the occurrence of a two-stage process that in fact corresponds to a cross-over behavior. Analyzing the spatial correlations with the measured time dependence of the MSD showed that the second process, yielding the increase of the MSD at longer times, is presumably associated with active centers in the spatial region that was formerly forming the interface between shear band and matrix during shear. Even though the time scales between experiment and simulation are not comparable, this observation is in excellent agreement with the conclusions drawn from radiotracer diffusion experiments [8] that gave rise to postulating the presence of interface-like states that could act as fast diffusion pathways, independent of the presence of dilated or densified regions along a given shear band. The fact that an enhanced MSD was observed long after the shear stress had been turned off substantiates the occurrence of a structural modification of the shear band regions serving to enhance mobility. Though an increase of the local excess volume was not explicitly found in MD simulation on accessible length scales, we still assume that it is the excess volume accumulation which promotes enhanced diffusion that is observed on a macro-scale. The generation of the excess volume might be inherently related to the stick-slip mode of plastic deformation [31] and an average dilatation related to a shear band (with alternating dense/dilute regions on a 100 nm scale) was unambiguously determined in dedicated TEM measurements [9, 10].

Since in the proposed scenario a shear band is initiated as an avalanche-like self-organization of STZs along a given quasi-2-dimensional path, one may expect that there is no total flux of excess volume from the outside; excess volume just redistributes along the shear band. Still, one may expect that at later stages of the shear band propagation some additional excess volume may migrate in from the outside (where a surface step is formed). This idea is in agreement with enhanced diffusion along a SB, see Ref. [8] and the present data, but apparently contradicts the STEM data [9, 10] that local denser areas are alternatively observed, too. Yet, such alternating dilatation/densification [9, 10] might be associated with a local topology of the shear band (which is not a plane) and the related in-plane compressive and tensile regions along the shear band path during propagation in a stick-slip manner [29], i.e. such regions might form at a later stage after the shear band has already been fully developed and serves as a slip "plane" with local deviations from the direction of maximum stress that act as local barriers for slip [10]. Thus, we cannot omit the hypothesis of excess volume migration into the shear band if we assume that this additional excess volume contributes to some 'delamination'-like events between the shear band and the surrounding matrix, leaving denser portions of the shear band unchanged with the concomitant formation of the above described 'interfaces' between the shear band and the matrix.

It is important to note that the existence of alternating denser and diluted regions provides conditions for mechanical (meta-)stability of the localized (extra) excess volume against macroscopic displacements of the matrix as a whole, as soon as diffusion events are frozen.

Adapting the volume-fluctuation model [32, 33] that has originally been developed for polycrystalline solids, the diffusion coefficient $D_{sb}$ of shear bands (in analogy to the diffusivity of grain boundaries) may be described as:

$$D_{sb} = D_v \exp\left(\frac{K_{sb} \Delta v_{sb}}{2 k_B T}\right) \qquad (1)$$

where $D_v$ is the bulk diffusion coefficient of the glass in the relaxed state, $K_{sb}$ is the bulk compressibility of the shear band, $\Delta v_{gb}$ is the deformation-induced excess volume per atom within the SB, and $k_B$ is Boltzmann's constant. The bulk compressibility of a shear band, $K_{sb}$, can be expressed as $K_{sb} = 2G_{sb}(1+v)/3(1-2v)$, where $G_{sb}$ is the shear modulus of the SB material and $v$ denotes Poisson's ratio. According to our direct measurements, only small changes of the elastic moduli and of Poisson's ratio were found in the deformed glass [34]. As a first-order approximation, we assume that the shear modulus of the shear band is similar to that of matrix.

The average additional excess volume of a shear band in PdNiP can be estimated as 0.4%, although local values can approach even 9 % [10]. Still, if we assume the formation of 'interfaces' between matrix and shear band, local values of $\Delta v_{gb}$ can safely be estimated as $0.1\Omega$, where $\Omega$ is the atomic volume, taken here as $10^{-29}$ m$^3$. Inserting this value into Eq. (2), assuming $G_{sb} = 50$ GPa and using $v = 0.4$ for PdNiP [34], the ratio $D_{sb}/D_v$ is about $6 \cdot 10^7$ at $T = 473$ K. Thus, Eq. (2) can reproduce the measured diffusion enhancement in shear bands over 7 to 8 orders of magnitude.

Let us analyze the relaxation behavior of the diffusion coefficient of a shear band. In the framework of the present model, it is the evolution of shear band compressibility $K_{sb}$ and that of the excess volume excess $\Delta v_{gb}$ which have to be evaluated.

Thermal annealing results in a non-monotonous evolution of the shear band diffusion coefficient. The behavior may be explained somehow similar to the idea of non-linear segregation suggested for the non-monotonous evolution of impurities at grain boundaries in crystalline materials [35, 36]. We suggest that there is first a re-organization of the additional excess volume in shear bands with a given time constant constrained by mechanical equilibrium induced by mechanical deformation, i.e. by the formed densified/diluted regions. This reorganization is accompanied by extra excess volume flow from neighboring matrix areas. As a result there is a further increase of the excess volume localized at shear bands and consequently diffusion enhances. According to Eq. (2), a 10% increase of the local excess volume would be sufficient to explain the results while formally keeping the compressibility unchanged. Since it could be expected that the compressibility decreases upon an increase of the excess volume, even smaller increases of the excess volume would be required. Further annealing results in relaxation of the accumulated excess volume and the disappearance of the dense/dilute contrast in SBs, which would also allow the glassy material inside the shear band to re-organize such that the highest-mobility pathways are annihilated, resulting in a decrease of the shear band diffusivity as observed in the experiments. This empirical model is fully in line with the observed cross-over behavior of the glass transition, since the characteristic time scales for the initial re-organization/re-distribution of excess volume and the relaxation of the accumulated excess volume inside the shear bands are expected to be different.

It is not only plastic deformation which modifies the local structure. Chemistry and the excess volume content of shear bands also trigger them into a different glassy state (without inducing nano-crystallization), but post-deformation relaxation imposes excess volume redistribution and the appearance of built-in residual stresses which affect the glass structure in terms of its corresponding potential energy landscape, thereby enabling the glass to trace new minima in the energy landscape. This behavior correlates with the fact that annealing does not erase glass memory on the previous plastic deformation which was found to persist even up to crystallization temperatures [37]. The present findings suggest a possibility of a further optimization of glass properties via combining deformation and heat treatments below the glass transition temperature. When we compare the energy input by thermal annealing and small deformation, we might not expect them to be very different. Yet, we see considerable changes, with the deformation being more efficient to alter the state and the further evolution/relaxation of the glass. This may be due to the "directionality" or due to the fact that deformation addresses the "right" sites (the ones that are most susceptible for shear). That might also be an issue concerning discussing the coupling of the Boson peak with the local structures and the energy landscape [37]. Moreover, time dependent stress relaxation along shear bands has recently been observed and a shear-band cavitation on micron scale has been reported [38]. Formation of cavities might be a finger-print of excess volume localization in addition to local tensile stresses, the existence of which was pointed out in Ref. [38].

The involved inherent relaxation dynamics of the shear band structure has a further consequence on the measured activation value of shear band diffusion, in fact the latter has to be treated with caution as an effective value, which is still characteristic for kinetic processes with the corresponding relaxation times.

In summary, we have shown that atomic diffusivity in shear bands is enhanced by 7 to 8 orders of magnitude with respect to the glassy matrix. The effective activation enthalpy of shear band diffusion is found to be about one-third of the corresponding value for diffusion in the glassy bulk (the undeformed matrix). Moreover, atomistic simulation supports directly the view that the zones of enhanced mobility (the short circuit diffusion paths) are localized at "interfaces" between shear bands and the glass matrix indicating an inherent inhomogeneity of the excess volume distribution upon deformation.

The relaxation kinetics of diffusion enhancement along shear bands is measured for the first time and an unexpected, strongly non-monotonous behaviour is found - the effective diffusion coefficient increases first and then decreases. These changes are going in line with the appearance of a distinct calorimetric "peak" at the glass transition temperature upon relaxation after deformation, which is attributed to inhomogeneous relaxation within the deformed glass. Moreover, a strong increase of the mean-squared displacement of atoms localized at a shear band is observed as a result of relaxation, substantiating that the evolution of the kinetics of atomic transport along shear bands is facilitated by excess volume re-distribution. In this respect the relaxation after deformation represents a localized "aging" effect. A phenomenological model of shear band initiation, evolution and relaxation is proposed.

**Methods**

**M1. Atomistic model and simulation details**

We apply a model for $Ni_{80}P_{20}$ which has been proposed by Kob and Andersen [39]. It is a binary mixture of Lennard-Jones (LJ) particles (say A and B) with ratio 80:20. This mixture is an archetypical model for a glass former. Particles interact via the LJ potential which is defined as:

$$U_{\alpha\beta}^{LJ}(r) = \varphi_{\alpha\beta}(r) - \varphi_{\alpha\beta}(r_c) - (r - r_c)\frac{d\varphi_{\alpha\beta}}{dr}(r_c), \varphi_{\alpha\beta}(r) = 4\varepsilon_{\alpha\beta}\left[\left(\frac{\sigma_{\alpha\beta}}{r}\right)^{12} - \left(\frac{\sigma_{\alpha\beta}}{r}\right)^{6}\right], \quad (M1)$$

for $r < r_c = 2.5\sigma_{AA}$ and zero otherwise (with $\alpha, \beta =$ A, B ). The interaction among the particles is defined as $\varepsilon_{AA} = 1.0$, $\varepsilon_{AB} = 1.5\varepsilon_{AA}$ and $\varepsilon_{BB} = 0.5\varepsilon_{AA}$. The range of interactions is given as $\sigma_{AA} = 1.0$, $\sigma_{AB} = 0.8\sigma_{AA}$ and $\sigma_{AB} = 0.88\sigma_{AA}$. The masses of both particles are equal, i.e., $m_A = m_B = 1.0$. All quantities are expressed in LJ units in which the unit of length is $\sigma_{AA}$, energy is expressed in the units of $\varepsilon_{AA}$, and the unit of time is $m_{AA}\sigma_{AA}^2/\varepsilon_{AA}$ (= $\tau_{LJ}$).

We introduce an elongated box geometry with dimensions $L_x \times L_y \times L_z = 20\sigma_{AA} \times 20\sigma_{AA} \times 80\sigma_{AA}$ and density 1.2. We consider 30,720 A-type particles and 7,680 B-type particles. The system is sheared at a constant shear rate using a simple planar Couette flow geometry, choosing *x* as the shear direction and *y* and *z* as the vorticity and gradient direction, respectively. Shear is imposed onto the system via Lees-Edwards boundary conditions [40] (modified periodic boundary conditions, where a particle that moves out of the simulation box in *z* direction is subject to a displacement in *x* direction according to the motion of the image cells below and above the simulation box with constant velocities -$U_{sx}$ and $U_{sx}$, respectively. With this scheme a linear velocity profile, $V_{s,x}(y) = \dot{\gamma}(y - \frac{L_z}{2})$ (with $\dot{\gamma} = U_{s,x}/L_z$ the shear rate), is obtained in the steady state.

**M2. Material and characterization**

PdNiP-based bulk metallic glass with the composition of $Pd_{40}Ni_{40}P_{20}$ (in at.%) was prepared by direct melting of palladium (purity 99.95 % ) and $Ni_2P$ powder (purity 99.5 % ) in an alumina crucible using an induction furnace in a purified argon atmosphere. The chemical compositions of the samples were confirmed by atomic absorption spectroscopy (Mikroanalytisches Labor Pascher, Germany). The crystalline master alloy was re-melted and then chill-cast into a copper mold with a 1×10×30 mm³ cavity.

A set of glassy samples was plastically deformed by cold rolling in one step at room temperature to various strains. The deformation degree ε was determined by the thickness reduction and the strain rate $\dot{\varepsilon}$ was about 5.5 s$^{-1}$. Cold rolling of the glassy samples led to shear band formation, which was detected by optical microscopy of the shear offset on the specimens surfaces. Plastic deformation to a relatively low strain of about 8 % was applied and the resulting shear band density, $\rho_{SB}$, was estimated at about 0.06 μm$^{-1}$.

X-ray diffraction was performed, using a Siemens D-5000 diffractometer equipped with a Cu cathode, a K$_\alpha$ monochromator, and a rotating sample holder in the θ–2θ geometry, using a point detector.

Differential scanning calorimetry (DSC) characterization was performed by a Perkin Elmer Diamond power compensated DSC applying a constant heating rate of 20 K/min in a temperature interval from 303 K to 798 K. The sample mass was about 20 - 30 mg. The measured signals on crystallized samples were used as base-line.

**M3. Radiotracer diffusion experiments**

The $^{110m}$Ag radioisotope (half-life of 252 days) with an initial specific activity of about 48 MBq/mg was produced by neutron irradiation of a natural silver chip at the research reactor GKSS, Geesthacht, Germany. The activated chip was first dissolved in 20 μl of HNO$_3$ and dissolved in 20 ml of double-distilled water. A droplet of the highly diluted $^{110m}$Ag solution was deposited onto the polished surface of each specimen and dried. The specimens were evacuated in silica ampoules to a residual pressure less than 10$^{-4}$ Pa, sealed, and annealed. The annealing temperatures were chosen well below the calorimetric glass transition temperature. After the annealing treatments, the samples were reduced in diameter to remove the effect of lateral and/or surface diffusion.

The penetration profiles were determined by the serial sectioning technique using a precision parallel grinder. As a key point we mention that a second γ-isotope, $^{59}$Fe, was further applied just before sectioning in order to check for the possible existence of micro-cracks (not seen by optical microscopy) and to guarantee the absence of any artifacts related to mechanical sectioning. The relative radioactivity of each section was measured with an intrinsic Ge γ-detector. The penetration profiles represent the plots of the measured relative specific radioactivity of the sections with subtracted background (which is proportional to the layer concentration of solute atoms) against the penetration depth, $y$, squared (according to the solution of the diffusion equation for the present boundary conditions). It was proven that the intensity of the $^{59}$Fe isotope decreases below the detection limit already after several sections, thereby validating the reliability of the penetration profiles measured for $^{110m}$Ag diffusion.

The relaxation behavior of shear bands at 498 K was investigated applying the radiotracer diffusion measurements as a sensitive probe of the shear band state. To this end, the diffusion sample, once annealed at 498 K for 3 days, was re-grinded and polished to background radioactivity, the $^{110m}$Ag was deposited again, and the diffusion annealing at 498 K was repeated for 3 more days. This step was repeated 3 times, so that the sample was finally annealed in total for 12 days. In parallel, DSC scans were performed after each heat treatment on a similar (non-radioactive) sample.

**References**

[1] Spaepen, F. Microscopic mechanism for steady-state inhomogeneous flow in metallic glasses. *Acta Metall*. **25**, 407- (1977).

[2] Argon, A. Plastic-deformation in metallic glasses. *Acta Metall*. **27**, 47-58 (1979).

**Acknowledgment**

The financial support of the Deutsche Forschungsgemeinschaft (DFG) in the framework of SPP1594 - Topological Engineering of Ultra-Strong Glasses is gratefully acknowledged.

**Author contributions**

I.B. conducted the experiments. G.P.S. and J. H. performed atomistic simulation. S.V.D. and G.W. developed the analytical model and performed the discussion and analysis of the experimental data. J.H. and G.W. initiated the work and supervised it at all stages. S.V.D. wrote the initial manuscript with input from all authors. All authors contributed to discussion of the results, provided input on the manuscript, and approved the final version.

**Additional information**

Supplementary information is available in the online version of the paper. Reprints and permissions information is available online at www.nature.com/reprints. Correspondence and requests for materials should be addressed to S.V.D. and G.W.

**Competing financial interests**

The authors declare no competing financial interests.